\documentclass[pra,aps]{revtex4}
\newcommand{\be}{\begin{equation}}
\newcommand{\ee}{\end{equation}}
\newcommand{\br}{\begin{eqnarray}}
\newcommand{\er}{\end{eqnarray}}
\newcommand{\nn}{\nonumber}
\newcommand{\bd}{\begin{displaymath}}
\newcommand{\ed}{\end{displaymath}}

\newcommand{\bib}{\bibitem}
\newcommand{\bfig}{\begin{figure}}
\newcommand{\efig}{\end{figure}}
\usepackage{ifvtex}
\usepackage{ifpdf}
\usepackage[dvips]{graphicx}
\usepackage{amsmath}
\usepackage{amssymb}
\usepackage{eucal}
\def\alf{\alpha}
\def\bet{\beta}

\def\om{\omega}
\def\eps{\epsilon}
\def\rpar{\right)}
\def\lpar{\left(}
\def\rbk{\right]}
\def\lbk{\left[}
\def\rbr{\right\}}
\def\lbr{\left\{}
\def\lb{\label}
\def\im{{\rm i}}
\def\tr{\mbox{${\rm Tr}$}}
\def\ro{\mbox{\boldmath $\rho$}}
\def\sig{\mbox{\boldmath $\sigma$}}
\def\ox{\mbox{$\textrm{\bf X}$}}
\def\ima{\mbox{${\rm Im}$}}
\def\re{\mbox{${\rm Re}$}}
\def\rg{\rangle}
\def\lg{\langle}

\begin{document}
%
\title{Extended Cahill-Glauber formalism for finite-dimensional spaces: \\
       II. Applications in quantum tomography and quantum teleportation}
\author{Marcelo A. Marchiolli}
\affiliation{Instituto de F\'{\i}sica de S\~{a}o Carlos, Universidade de S\~{a}o Paulo, \\
             Caixa Postal 369, 13560-970, S\~{a}o Carlos, SP, Brazil \\
             E-mail address: marcelo$\_$march@bol.com.br}
\author{Maurizio Ruzzi, Di\'{o}genes Galetti}
\affiliation{Instituto de F\'{\i}sica Te\'{o}rica, Universidade Estadual Paulista, \\
             Rua Pamplona 145, 01405-900, S\~{a}o Paulo, SP, Brazil \\
             E-mail address: mruzzi@ift.unesp.br and galetti@ift.unesp.br}
\date{\today}
%
\begin{abstract}
\vspace*{0.1mm}
\begin{center}
\rule[0.1in]{142mm}{0.4mm}
\end{center}
By means of a new mod$(N)$ invariant operator basis, $s$-parametrized phase-space functions associated with bounded operators in a
finite-dimensional Hilbert space are introduced in the context of the extended Cahill-Glauber formalism, and their properties are
discussed in details. The discrete Glauber-Sudarshan, Wigner, and Husimi functions emerge from this formalism as specific cases of
$s$-parametrized phase-space functions where, in particular, a hierarchical process among them is promptly established. In addition, a
phase-space description of quantum tomography and quantum teleportation is presented and new results are obtained. \\
\vspace*{0.1mm}
\begin{center}
\rule[0.1in]{142mm}{0.4mm}
\end{center}
\end{abstract}
\maketitle
\section{Introduction}

The first proposal of a unified formalism for quasiprobability distribution functions in continuous phase space has its origin in the
seminal works produced by Cahill and Glauber \cite{r1}. Since then a huge number of papers have appeared in the literature covering a
wide range of practical applications in different physical systems modeled by means of infinite-dimensional Hilbert spaces \cite{r2,r3}.
In particular, the phase-space description of some important effects in quantum mechanics, such as interference, entanglement, and 
decoherence, has opened up astounding possibilities for the comprehension of intriguing aspects of the microscopic world \cite{r4}.
However, if physical systems with a finite-dimensional space of states are considered, then the quasiprobability distribution functions
are described by a set of discrete variables defined over a finite lattice \cite{r5,r6,r7,r8,r9,r10,r11,r12,r13,r14,r15}. In this sense,
Opatrn\'{y} {\em et al} \cite{r9} were the first researchers to propose a unified approach to the problem of discrete quasiprobability
distribution functions in the literature. Basically, they used a discrete displacement-operator expansion to introduce $s$-parametrized
phase-space functions associated with operators defined over a finite-dimensional Hilbert space. Furthermore, the authors showed that
the discrete Glauber-Sudarshan, Wigner, and Husimi functions are particular cases of $s$-parametrized phase-space functions and depend
on the arbitrary reference state whose characteristic function cannot have zero values. It is worth mentioning that the dependence on
the right choice of the reference state and the associated problems with the mod$(N)$ invariance of the discrete displacement operators
represent two important restrictions inherent to their approach which deserve to be carefully investigated. Nowadays, beyond these
fundamental features, discrete quasiprobability distribution functions in finite-dimensional phase spaces have potential applications
for quantum-state tomography \cite{r16,r17}, quantum teleportation \cite{r18,r19,r20,r21}, phase-space representation of quantum
computers \cite{r22}, open quantum systems \cite{r23}, quantum information theory \cite{r24}, and quantum computation \cite{r25}.

The main aim of this paper is to present a consistent formalism for the quasiprobability distribution functions defined over a discrete
$N^{2}$-dimensional phase space, which is based upon the mathematical fundamentals developed in \cite{r26}. First, we review important
topics and introduce new properties concerning the mod$(N)$ invariant operator basis which leads us not only to define a parametrized
phase-space function in terms of the discrete $s$-ordered characteristic function, but also to discuss some characteristics inherent to
the extended Cahill-Glauber formalism for finite-dimensional spaces. The restriction on the right choice of the reference state is
overcome in this approach through the vacuum state established by Galetti and de Toledo Piza \cite{r8}, whose analytical properties were
extensively explored in \cite{r10}. Consequently, the discrete Glauber-Sudarshan, Wigner, and Husimi functions are well-defined in the
present context and represent specific cases of $s$-parametrized phase-space functions describing density operators associated with
physical systems whose space of states is finite. In addition, we also establish a hierarchical order among them through a smoothing
process characterized by a discrete phase-space function that closely resembles the role of a Gaussian function in the continuous
phase-space. In this point, it is worth emphasizing that our {\it ab initio} construction inherently embodies the discrete analogues of
the desired properties of the Cahill-Glauber approach. Next, we apply such discrete extension into the context of quantum information
processing, quantum tomography, and quantum teleportation in order to obtain a phase-space description of some topics related to unitary
depolarizers, discrete Radon transforms, and generalized Bell states. In particular, we attain new results within which some of them
deserve to be mentioned: (i) we show that the symmetrized Schwinger operator basis introduced in \cite{r6} can be considered a unitary
depolarizer; (ii) we establish a link between measurable quantities and $s$-ordered characteristic functions by means of discrete Radon
transforms, which can be used to construct any quasiprobability distribution functions defined over a $N^{2}$-dimensional phase space;
and finally, (iii) we present a quantum teleportation protocol that leads us to reach a generalized phase-space description of the
physical process discussed by Bennett {\em et al} \cite{r18}.

This paper is organized as follows. In section II we present some basic properties inherent to the new discrete mapping kernel which
allow us to define a parametrized phase-space function in terms of a discrete $s$-ordered characteristic function. Following, in section
III we show that the extended Cahill-Glauber formalism not only introduces new mathematical tools for the analysis of finite quantum
systems, but also can be applied in the context of quantum information processing, quantum tomography, and quantum teleportation.
Moreover, we also employ a slightly modified version of the scattering circuit to measure any discrete Wigner function in the
phase-space representation. Finally, section IV contains our summary and conclusions.
 
\section{The mapping kernel}

There is a huge variety of probability distribution functions defined in continuous quantum phase-spaces whose range of practical
applications in physics covers different areas and scenarios \cite{r2,r3}. For example, the well-known Cahill-Glauber formalism
\cite{r1} provides a general mapping technique of bounded operators which permits, in particular, to define a generalized probability
distribution function $F^{(s)}(q,p) = \tr [ {\bf T}^{(s)}(q,p) \ro ]$ associated with an arbitrary physical system described by the
density operator $\ro$. In this approach, the mapping kernel (hereafter $\hbar = 1$)
\be
\lb{s2e1}
{\bf T}^{(s)}(q,p) = \int \frac{dq^{\prime} dp^{\prime}}{2 \pi} \exp [ \im ( q^{\prime} p - p^{\prime} q ) ] \, {\bf D}^{(s)}
(q^{\prime},p^{\prime})
\ee
is defined as a Fourier transform of the parametrized operator
\be
\lb{s2e2}
{\bf D}^{(s)}(q^{\prime},p^{\prime}) = \exp [ (s/4) ( q^{\prime 2} + p^{\prime 2} ) ] \, {\bf D}(q^{\prime},p^{\prime})
\ee
where ${\bf D}(q^{\prime},p^{\prime}) = \exp [ \im (p^{\prime} {\bf Q} - q^{\prime} {\bf P}) ]$ is the usual displacement operator 
written in terms of the coordinate and momentum operators satisfying the Weyl-Heisenberg commutation relation $[ {\bf Q},{\bf P} ] = 
\im {\bf 1}$, and $s$ is a complex parameter. Thus, for $s =-1,0,+1$ the generalized probability distribution function leads to the
so-called Husimi, Wigner and Glauber-Sudarshan functions, respectively. Besides, these functions present specific properties and
correspond to different ordered power-series expansions in the annihilation and creation operators of the density operator: the Husimi
function $\mathcal{H}(q,p)$ is infinitely differentiable and it is associated with the normally ordered form; the Wigner function
$\mathcal{W}(q,p)$ is a continuous and uniformly bounded function, it can take negative values and corresponds to the symmetrically
ordered form; and finally, the Glauber-Sudarshan function $\mathcal{P}(q,p)$ is highly singular; it does not exist as a regular function
for pure states and it corresponds to the antinormally ordered form. After this condensed review of the Cahill-Glauber formalism for the
quasiprobability distribution functions, we will establish the discrete representatives of these functions in an $N^{2}$-dimensional
phase space.

\subsection{The new $\textrm{mod}(N)$ invariant operator basis}

Let us introduce the symmetrized version of the unitary operator basis proposed by Schwinger \cite{r27} as
\be
\lb{s2e3}
{\bf S}(\eta,\xi) = \frac{1}{\sqrt{N}} \exp \lpar \frac{\im \pi}{N} \eta \xi \rpar {\bf U}^{\eta} {\bf V}^{\xi}
\ee
where the labels $\eta$ and $\xi$ are associated with the dual coordinate and momentum variables of a discrete $N^{2}$-dimensional
phase space. Consequently, these labels assume integer values in the symmetrical interval $[-\ell,\ell]$, with $\ell = (N-1)/2$. A
comprehensive and useful compilation of results and properties of the unitary operators ${\bf U}$ and ${\bf V}$ can be found in 
reference \cite{r10}, since the initial focus of our attention is the essential features exhibited by (\ref{s2e3}). Note that the set 
of $N^{2}$ operators $\{ {\bf S}(\eta,\xi) \}_{\eta,\xi = -\ell, \ldots, \ell}$ constitutes a complete orthonormal operator basis
which allows us, in principle, to construct all possible dynamical quantities belonging to the system \cite{r27}. Thus, the 
decomposition of any linear operator ${\bf O}$ in this basis is written as
\be
\lb{s2e4}
{\bf O} = \sum_{\eta,\xi = - \ell}^{\ell} \mathcal{O}(\eta,\xi) {\bf S}(\eta,\xi)
\ee
with the coefficients $\mathcal{O}(\eta,\xi)$ given by $\tr [ {\bf S}^{\dagger}(\eta,\xi) {\bf O} ]$. It must be stressed that this
decomposition is unique since the relations ${\bf S}^{\dagger}(\eta,\xi) = {\bf S}(-\eta,-\xi)$ and $\tr [ {\bf S}^{\dagger}(\eta,\xi)
{\bf S}(\eta^{\prime},\xi^{\prime}) ] = \delta_{\eta^{\prime},\eta}^{[N]} \delta_{\xi^{\prime},\xi}^{[N]}$ are promptly verified. The
superscript $[N]$ on the Kronecker delta denotes that this function is different from zero when its labels are ${\rm mod}(N)$ congruent.

The new ${\rm mod}(N)$ invariant operator basis recently proposed in \cite{r26},
\be
\lb{s2e5}
{\bf T}^{(s)}(\mu,\nu) = \frac{1}{\sqrt{N}} \sum_{\eta,\xi = -\ell}^{\ell} \exp \lbk \im \pi \Phi(\eta,\xi;N) - \frac{2 \pi \im}{N}
(\eta \mu + \xi \nu) \rbk {\bf S}^{(s)}(\eta,\xi) \; ,
\ee
is defined by means of a discrete Fourier transform of the extended mapping kernel
\bd
{\bf S}^{(s)}(\eta,\xi) = \lbk \mathcal{K}(\eta,\xi) \rbk^{-s} {\bf S}(\eta,\xi)
\ed
where the extra term $\mathcal{K}(\eta,\xi)$ can be expressed as a sum of products of Jacobi theta functions evaluated at integer
arguments \cite{r28},
\br
\lb{s2e6}
\mathcal{K}(\eta,\xi) &=& \lbr 2 \lbk \vartheta_{3} (0|\im a) \vartheta_{3}(0|4 \im a) + \vartheta_{4}(0|\im a) \vartheta_{2}(0|4 \im a)
\rbk \rbr^{-1} \lbr \vartheta_{3}(\pi a \eta | \im a) \vartheta_{3}(\pi a \xi | \im a) + \vartheta_{3}(\pi a \eta | \im a) \vartheta_{4}
(\pi a \xi | \im a) \exp (\im \pi \eta) \right. \nn \\
& & + \left. \vartheta_{4}(\pi a \eta | \im a) \vartheta_{3}(\pi a \xi | \im a) \exp (\im \pi \xi) + \vartheta_{4}(\pi a \eta | \im a)
\vartheta_{4}(\pi a \xi | \im a) \exp [\im \pi (\eta + \xi + N) ] \rbr 
\er
with $a = (2N)^{-1}$. As mentioned in \cite{r26}, $\mathcal{K}(\eta,\xi)$ is a bell-shaped function in the discrete variables
$(\eta,\xi)$ and equals to one for $\eta = \xi = 0$; in addition, the complex parameter $s$ obeys $| s | \leq 1$. The phase
$\Phi(\eta,\xi;N) = N \textrm{I}_{\eta}^{N} \textrm{I}_{\xi}^{N} - \eta \textrm{I}_{\xi}^{N} - \xi \textrm{I}_{\eta}^{N}$ is responsible
for the $\textrm{mod}(N)$ invariance of the operator basis (\ref{s2e5}), $\textrm{I}_{\sigma}^{N} = [ \sigma/N ]$ being the integral
part of $\sigma$ with respect to $N$. This definition stands for the discrete version of the continuous mapping kernel (\ref{s2e1}) and
represents the cornerstone of the present approach.

By analogy with decomposition (\ref{s2e4}), the expansion
\be
\lb{s2e7}
{\bf O} = \frac{1}{N} \sum_{\mu,\nu = - \ell}^{\ell} \mathcal{O}^{(-s)}(\mu,\nu) {\bf T}^{(s)}(\mu,\nu)
\ee
can also be verified for any linear operator. Here, the coefficients $\mathcal{O}^{(-s)}(\mu,\nu) = \tr [ {\bf T}^{(-s)}(\mu,\nu) 
{\bf O} ]$ correspond to a one-to-one mapping between operators and functions belonging to an $N^{2}$-dimensional phase space 
characterized by the discrete labels $\mu$ and $\nu$. In particular, if one considers $s=-1$ and ${\bf O} = \ro$ in equation
(\ref{s2e7}), we obtain the diagonal representation
\be
\lb{s2e8}
\ro = \frac{1}{N} \sum_{\mu,\nu=-\ell}^{\ell} \mathcal{P}(\mu,\nu) | \mu,\nu \rangle \langle \mu,\nu |
\ee
where $\mathcal{P}(\mu,\nu) = \tr \lbk {\bf T}^{(1)}(\mu,\nu) \ro \rbk$ is the discrete version of the Glauber-Sudarshan function for
finite Hilbert spaces, and ${\bf T}^{(-1)}(\mu,\nu)$ is the projector of discrete coherent-states \cite{r10}. For $s=0$, we verify that
\be
\lb{s2e9}
\ro = \frac{1}{N} \sum_{\mu,\nu=-\ell}^{\ell} \mathcal{W}(\mu,\nu) {\bf G}(\mu,\nu)
\ee
recovers the well-established results in \cite{r8}, $\mathcal{W}(\mu,\nu) = \tr [ {\bf G}^{\dagger}(\mu,\nu) \ro ]$ being the discrete
Wigner function and ${\bf G}(\mu,\nu)$ the $\textrm{mod}(N)$ invariant operator basis whose mathematical properties were studied in
\cite{r10}. Furthermore, we note that the Husimi function in the discrete coherent state representation, $\mathcal{H} (\mu,\nu) = \tr
[ {\bf T}^{(-1)}(\mu,\nu) \ro ]$, can be promptly obtained from equations (\ref{s2e8}) or (\ref{s2e9}) by means of a trace operation.
Next, we will discuss some properties inherent to the set of $N^{2}$ operators $\{ {\bf T}^{(s)}(\mu,\nu) \}_{\mu , \nu = -\ell, \ldots,
\ell}$ with emphasis on establishing a hierarchical process among the quasiprobability distribution functions in finite-dimensional
spaces.

\subsection{Basic properties}

The discrete mapping kernel ${\bf T}^{(s)}(\mu,\nu)$ presents some inherent mathematical features that lead us to derive a set of
properties which characterize its algebraic structure. For instance, it is straightforward to show that the equalities 
\br
\textrm{(i)} &\! \! \!& \frac{1}{N} \sum_{\mu,\nu = - \ell}^{\ell} {\bf T}^{(s)}(\mu,\nu) = {\bf 1} \nn \\
\textrm{(ii)} &\! \! \!& \tr \lbk {\bf T}^{(s)}(\mu,\nu) \rbk = 1 \nn \\
\textrm{(iii)} &\! \! \!& \tr \lbk {\bf T}^{(-s)}(\mu,\nu) {\bf T}^{(s)}(\mu^{\prime},\nu^{\prime}) \rbk = N
\delta_{\mu^{\prime},\mu}^{[N]} \delta_{\nu^{\prime},\nu}^{[N]} \nn
\er
are promptly verified where, in particular, the third property has been reached with the help of the auxiliary relation 
\bd
\tr \lbk {\bf T}^{(t)}(\mu,\nu) {\bf T}^{(s)}(\mu^{\prime},\nu^{\prime}) \rbk = \frac{1}{N} \sum_{\eta,\xi=-\ell}^{\ell} \exp \lbr
\frac{2 \pi \im}{N} \lbk \eta \lpar \mu^{\prime} - \mu \rpar + \xi \lpar \nu^{\prime} - \nu \rpar \rbk \rbr \lbk \mathcal{K}(\eta,\xi)
\rbk^{-(t + s)} \; .
\ed
Note that for $s=-1$, the first property coincides with the completeness relation of the discrete coherent states (the proof of this
relation was given in \cite{r10}); the second property simply states that ${\bf T}^{(s)}(\mu,\nu)$ has a unit trace. Finally, the
third property is the counterpart to the orthogonality rule established for the operators ${\bf S}(\eta,\xi)$. Furthermore, we also
verify the condition ${\bf T}^{(s^{\ast})}(\mu,\nu) = [ {\bf T}^{(s)}(\mu,\nu) ]^{\dagger}$, which implies that for real values of the
parameter $s$, the discrete mapping kernel is Hermitian; consequently, the mappings of Hermitian operators in the $N^{2}$-dimensional
phase space lead us to obtain real functions. Now, let us establish a hierarchical process among the discrete Glauber-Sudarshan, Wigner
and Husimi functions.

The connection between the discrete Glauber-Sudarshan and Wigner functions is reached with the help of equation (\ref{s2e8}) through a
smoothing process of ${\cal P}(\mu,\nu)$, i.e., 
\be
\lb{s2e10}
\mathcal{W}(\mu,\nu) = \frac{1}{N} \sum_{\mu^{\prime},\nu^{\prime} = - \ell}^{\ell} \textrm{E}(\mu^{\prime} - \mu,\nu^{\prime} - \nu)
\mathcal{P}(\mu^{\prime},\nu^{\prime})
\ee
where $\textrm{E}(\mu^{\prime}-\mu,\nu^{\prime}-\nu) \equiv \tr [{\bf T}^{(0)}(\mu,\nu) {\bf T}^{(-1)}(\mu^{\prime},\nu^{\prime})]$ is 
expressed by means of a discrete Fourier transform of the function $\mathcal{K}(\eta,\xi)$ -- note that $\textrm{E} (\mu^{\prime}-\mu,
\nu^{\prime}-\nu)$ can be interpreted as a Wigner function evaluated for the discrete coherent states labeled by $\mu^{\prime}$ and
$\nu^{\prime}$. Similarly, the link between discrete Wigner and Husimi functions can also be established through equation (\ref{s2e9})
as follows:
\be
\lb{s2e11}
\mathcal{H}(\mu,\nu) = \frac{1}{N} \sum_{\mu^{\prime},\nu^{\prime} = -\ell}^{\ell} \textrm{E}(\mu^{\prime} - \mu,\nu^{\prime} - \nu) 
\mathcal{W}(\mu^{\prime},\nu^{\prime}) \; .
\ee
Therefore, equations (\ref{s2e10}) and (\ref{s2e11}) exhibit a sequential smoothing which characterizes a hierarchical process among the
quasiprobability distribution functions in finite-dimensional spaces, $\mathcal{P}(\mu,\nu) \rightarrow \mathcal{W}(\mu,\nu) \rightarrow
\mathcal{H}(\mu,\nu)$. It is worth mentioning that
\be
\lb{s2e12}
\mathcal{H}(\mu,\nu) = \frac{1}{N} \sum_{\mu^{\prime},\nu^{\prime} = -\ell}^{\ell} \left| \lg \mu,\nu | \mu^{\prime},\nu^{\prime} \rg
\right|^{2} \mathcal{P}(\mu^{\prime},\nu^{\prime})
\ee
establishes an additional relation which allows us to connect both the discrete Husimi and Glauber-Sudarshan functions without the
intermediate process given by $\mathcal{W}(\mu,\nu)$, being $| \lg \mu,\nu | \mu^{\prime},\nu^{\prime} \rg |^{2} = | \mathcal{K}
(\mu^{\prime} - \mu,\nu^{\prime} - \nu) |^{2}$ the overlap probability for discrete coherent states. Opatrn\'{y} {\em et al} \cite{r9}
have used a similar formalism in order to establish a set of parametrized discrete phase-space functions for finite-dimensional Hilbert
spaces, where some mathematical procedures were introduced to circumvent the condition of mod$(N)$ invariance of the discrete
displacement operators. In that approach, the discrete $s$-parametrized functions basically depend on the arbitrary reference state
whose characteristic function cannot have zero values. Here, we have established a suitable mathematical procedure that allows us to
overcome some intrinsic problems encountered in \cite{r9}, being the vacuum state defined in \cite{r8,r10} as our reference state.

Next, we present two important properties associated with the trace of the product of two bounded operators and the matrix elements
$\lg m | {\bf T}^{(s)}(\mu,\nu) | n \rg$ in the finite number basis $\{ | n \rg \}_{n=0,\ldots,N-1}$. The first one corresponds to the
overlap
\bd
\textrm{(iv)} \; \; \tr ({\bf A} {\bf B}) = \frac{1}{N} \sum_{\mu,\nu = - \ell}^{\ell} \mathcal{A}^{(-s)}(\mu,\nu) \mathcal{B}^{(s)}
(\mu,\nu) 
\ed
where, in particular, for $s=0$, the trace of the product of two density operators coincides with the overlap of the discrete Wigner
functions of each density operator,
\bd
\tr (\ro_{1} \ro_{2}) = \frac{1}{N} \sum_{\mu,\nu = - \ell}^{\ell} \mathcal{W}_{1}(\mu,\nu) \mathcal{W}_{2}(\mu,\nu) \; .
\ed
In addition, the mean value of any bounded operator can also be obtained from this property,
\be
\lb{s2e13}
\lg {\bf O} \rg \equiv \tr ({\bf O} \ro) = \frac{1}{N} \sum_{\mu,\nu = - \ell}^{\ell} \mathcal{O}^{(-s)}(\mu,\nu) F^{(s)}(\mu,\nu)
\ee
being the parametrized function $F^{(s)}(\mu,\nu)$ defined as the expectation value of the discrete mapping kernel (\ref{s2e5}), i.e.,
\be
\lb{s2e14}
F^{(s)}(\mu,\nu) \equiv \tr \lbk {\bf T}^{(s)}(\mu,\nu) \ro \rbk = \frac{1}{\sqrt{N}} \sum_{\eta,\xi = -\ell}^{\ell} \exp \lbk \im \pi
\Phi(\eta,\xi;N) - \frac{2 \pi \im}{N} (\eta \mu + \xi \nu) \rbk \Xi^{(s)}(\eta,\xi)
\ee
while $\Xi^{(s)}(\eta,\xi) \equiv \tr [ {\bf S}^{(s)}(\eta,\xi) \ro ]$ represents the discrete $s$-ordered characteristic function 
\cite{r1}. Note that $\Phi(\eta,\xi;N)$ can be discarded in equation (\ref{s2e14}) since the discrete labels $\eta$ and $\xi$ are
confined into the closed interval $[ -\ell,\ell ]$. In fact, this phase will be important only in the mapping of the product of $M$
quantum operators \cite{r10}. Besides, for $s=-1,0,+1$ the parametrized function is directly related to the discrete Husimi, Wigner and
Glauber-Sudarshan functions, respectively. Hence, the characteristic function can now be promptly calculated for each situation through
the inverse discrete Fourier transform of the generalized probability distribution function $F^{(s)}(\mu,\nu)$.

The second one refers to the nondiagonal matrix elements in the finite number basis
\bd
{\rm (v)} \; \; \lg m | {\bf T}^{(s)}(\mu,\nu) | n \rg = \frac{1}{N} \sum_{\eta,\xi = -\ell}^{\ell} \exp \lbk \im \pi \Phi(\eta,\xi;N) -
\frac{2 \pi \im}{N} (\eta \mu + \xi \nu) \rbk [ {\cal K}(\eta,\xi) ]^{-s} \Gamma_{mn} (\eta,\xi)
\ed
with
\be
\lb{s2e15}
\Gamma_{mn}(\eta,\xi) = \exp \lpar - \frac{\im \pi}{N} \eta \xi \rpar \sum_{\sigma=-\ell}^{\ell} \exp \lpar \frac{2 \pi \im}{N} \sigma
\eta \rpar \mathfrak{F}_{\sigma,n} \, \mathfrak{F}_{\sigma - \xi,m}^{\ast}
\ee
written in terms of the coefficients \cite{r8}
\bd
\mathfrak{F}_{\kappa,n} = \textrm{N}_{n} \frac{(-\im)^{n}}{\sqrt{N}} \sum_{\beta = -\infty}^{\infty} \exp \lpar - \frac{\pi}{N} 
\beta^{2} + \frac{2 \pi \im}{N} \beta \kappa \rpar \textrm{H}_{n} \lpar \sqrt{\frac{2 \pi}{N}} \, \beta \rpar 
\ed
where $\textrm{N}_{n}$ is the normalization constant, and $\textrm{H}_{n}(z)$ is a Hermite polynomial. It is easy to show that
$\Gamma_{mn}(\eta,\xi)$ satisfies the relations $\Gamma_{mn}(0,0)=\delta_{m,n}^{[N]}$ and $\Gamma_{00}(\eta,\xi)=\mathcal{K}(\eta,\xi)$,
which are associated with the orthogonality rule for the finite number states and the diagonal matrix element $\lg 0 | {\bf T}^{(s)}
(\mu,\nu) | 0 \rg$ for the vacuum state. Moreover, adopting the mathematical procedure established in \cite{r29} for the continuum
limit, we obtain
\bd
\Gamma_{mn}(q^{\prime},p^{\prime}) = \sqrt{\frac{m!}{n!}} \lpar \frac{q^{\prime} + \im p^{\prime}}{\sqrt{2}} \rpar^{n-m} 
\textrm{L}_{m}^{(n-m)} \lpar \frac{| q^{\prime} + \im p^{\prime} |^{2}}{2} \rpar \exp \lbk - \frac{1}{4} (q^{\prime 2} + p^{\prime 2}) 
\rbk \qquad (n \geq m)
\ed
with $\textrm{L}_{n}^{(m)}(z)$ being the associated Laguerre polynomial. Consequently, the nondiagonal matrix elements for $|s|<1$
take the analytical form
\bd
\lg m | {\bf T}^{(s)}(q,p) | n \rg = \frac{2}{1-s} \, \sqrt{\frac{m!}{n!}} \lpar - \frac{1+s}{1-s} \rpar^{m} \lbk \frac{\sqrt{2} 
(q- \im p)}{1-s} \rbk^{n-m} \textrm{L}_{m}^{(n-m)} \lbk \frac{2 (q^{2}+p^{2})}{1-s^{2}} \rbk \exp \lpar - \frac{q^{2}+p^{2}}{1-s}
\rpar \; .
\ed
This result coincides exactly with that obtained by Cahill and Glauber \cite{r1} for the mapping kernel (\ref{s2e1}), since 
${\bf T}^{(s)}(\mu,\nu)$ goes to ${\bf T}^{(s)}(q,p)$ in the limit $N \rightarrow \infty$. Following, we will discuss some applications
for the generalized probability distribution function $F^{(s)}(\mu,\nu)$ with emphasis on the discrete phase-space representation of
quantum tomography and quantum teleportation.

\section{Applications}

Nowadays, within the context of quasiprobability distribution functions in finite-dimensional spaces, the discrete Wigner function has a
central role in some recent researches on quantum-state tomography \cite{r16,r17}, quantum teleportation \cite{r18,r19,r20,r21},
phase-space representation of quantum computers \cite{r22}, open quantum systems \cite{r23}, quantum information theory \cite{r24},
and quantum computation \cite{r25}. Basically, these works are based on the well-established Wootters' approach \cite{r5} for discrete
Wigner functions, in which ``the field of real numbers that labels the axes of continuous phase space is replaced by a finite field
having $N$ elements," $N$ being the power of a prime number. Notwithstanding this, there are other formalisms for finite-dimensional
Hilbert spaces with convenient inherent mathematical properties which can also be applied in the description of similar quantum systems
\cite{r6,r7,r8,r9,r10,r11,r12,r13,r14,r15,r26}. In this section, we will show that the present formalism not only introduces new
mathematical tools for the analysis of finite quantum systems but also can be applied, for example, to the context of quantum
information processing, quantum tomography and quantum teleportation.

\subsection{Quantum information processing}

Within the most important quantum operations in quantum information processing, unitary operations have a prominent position \cite{r30}.
Besides, in the scope of quantum information theory, the unitary depolarizers play an important role in quantum teleportation and
quantum dense coding \cite{r21,r31}. With respect to $N$-dimensional Hilbert spaces, unitary depolarizers are defined on a domain
$\Omega$ as elements of the set
\be
\lb{s3e1}
\mathcal{D}(N) = \lbr \ox_{\epsilon} \left| \; \ox_{\epsilon} \ox_{\epsilon}^{\dagger} \right. = \ox_{\epsilon}^{\dagger} \ox_{\epsilon}
= {\bf 1}, \; \epsilon \in \Omega \rbr
\ee
which satisfy the relation
\be
\lb{s3e2}
\frac{1}{N} \sum_{\epsilon \in \Omega} \ox_{\epsilon} {\bf O} \ox_{\epsilon}^{\dagger} = \tr ({\bf O}) {\bf 1}
\ee
for any linear operator ${\bf O}$ acting on finite-dimensional vector spaces, where ${\bf 1}$ is an identity operator. Recently, Ban
\cite{r32} has shown that the Pegg-Barnett phase operator formalism is useful for quantum information processing as well as in
investigating quantum optical systems. In this sense, it is worth mentioning that the symmetrized version of the Schwinger operator
basis ${\bf S}(\eta,\xi)$ can also be considered a unitary depolarizer, since the elements of the set
\be
\lb{s3e3}
\mathcal{D}(N) = \lbr \sqrt{N} {\bf S}(\eta,\xi) \left| \; [ \sqrt{N} {\bf S}(\eta,\xi) ] [ \sqrt{N} {\bf S}(\eta,\xi) ]^{\dagger}
\right. = [ \sqrt{N} {\bf S}(\eta,\xi) ]^{\dagger} [ \sqrt{N} {\bf S}(\eta,\xi) ] = {\bf 1}, \; - \ell \leq \eta , \xi \leq \ell \rbr
\ee
obey the property
\be
\lb{s3e4}
\frac{1}{N} \sum_{\eta,\xi = -\ell}^{\ell} [ \sqrt{N} {\bf S}(\eta,\xi) ] {\bf O} [ \sqrt{N} {\bf S}(\eta,\xi) ]^{\dagger} = \tr 
({\bf O}) {\bf 1} \; .
\ee
This result shows that the average over all possible discrete dual coordinate and momentum shifts on the $N^{2}$-dimensional phase
space completely randomizes any quantum state defined on the finite-dimensional vector space. Furthermore, for $s = \im \omega$ and
$\omega \in \mathbb{R}$, the elements of the set $\{ \sqrt{N} {\bf S}^{(\im \omega)}(\eta,\xi) \}_{\eta,\xi = -\ell, \ldots, \ell}$
generalize equation (\ref{s3e3}), being ${\bf S}^{(\im \omega)}(\eta,\xi)$ the parametrized Schwinger operator basis. Unfortunately,
the implementation of such unitary operations in a realistic quantum-computer technology encounters an almost unsurmountable obstacle:
the degrading and ubiquitous decoherence due to the unavoidable coupling with the environment \cite{r33}. However, recent progress
\cite{r34} has developed the idea of protecting or even creating a decoherence-free subspace for processing quantum information.

\subsection{Marginal distributions, Radon transforms and discrete phase-space tomography}

The marginal distributions associated with the generalized probability distribution function $F^{(s)}(\mu,\nu)$ are obtained through the
usual mathematical procedure
\br
\lb{s3e5}
\mathcal{Q}^{(s)}(\mu) &\equiv& \frac{1}{\sqrt{N}} \sum_{\nu = - \ell}^{\ell} F^{(s)}(\mu,\nu) = \sum_{\eta = - \ell}^{\ell} \exp \lpar
- \frac{2 \pi \im}{N} \eta \mu \rpar \Xi^{(s)}(\eta,0) \\
\lb{s3e6}
\mathcal{R}^{(s)}(\nu) &\equiv& \frac{1}{\sqrt{N}} \sum_{\mu = - \ell}^{\ell} F^{(s)}(\mu,\nu) = \sum_{\xi = - \ell}^{\ell} \exp \lpar -
\frac{2 \pi \im}{N} \xi \nu \rpar \Xi^{(s)}(0,\xi) \; .
\er
Note that the second equality in both definitions has been attained with the help of equation (\ref{s2e14}). Consequently, the marginal
distributions are obtained by means of discrete Fourier transforms of the $s$-ordered characteristic function calculated in specific
slices of the dual plane $(\eta,\xi)$. Now, if one considers the hierarchical process established by equations (\ref{s2e10}) and
(\ref{s2e11}), alternative expressions for the marginal distributions associated with the Wigner and Husimi functions can also be 
derived,
\br
\mathcal{Q}^{(0)}(\mu) = \sum_{\mu^{\prime} = - \ell}^{\ell} \mathcal{E}(\mu^{\prime} - \mu) \mathcal{Q}^{(1)}(\mu^{\prime}) \qquad
\qquad \mathcal{R}^{(0)}(\nu) = \sum_{\nu^{\prime} = - \ell}^{\ell} \mathcal{E}(\nu^{\prime} - \nu) \mathcal{R}^{(1)}(\nu^{\prime}) 
\nn \\
\mathcal{Q}^{(-1)}(\mu) = \sum_{\mu^{\prime} = - \ell}^{\ell} \mathcal{E}(\mu^{\prime} - \mu) \mathcal{Q}^{(0)}(\mu^{\prime}) \qquad
\qquad \mathcal{R}^{(-1)}(\nu) = \sum_{\nu^{\prime} = - \ell}^{\ell} \mathcal{E}(\nu^{\prime} - \nu) \mathcal{R}^{(0)}(\nu^{\prime}) \nn 
\er
where the smoothing function $\mathcal{E}(\chi)$ is given by
\bd
\mathcal{E}(\chi) = \dfrac{1}{\sqrt{2N}} \, \dfrac{\vartheta_{3}(0|\im a) \vartheta_{3}(2 \pi a \chi |\im a) + \vartheta_{4}(0|\im a)
\vartheta_{4}(2 \pi a \chi |\im a)} {\vartheta_{3}(0|\im a) \vartheta_{3}(0|4 \im a) + \vartheta_{4}(0|\im a) \vartheta_{2}(0|4 \im a)}
\; .
\ed
Thus, a sequential smoothing process is immediately established among the discrete marginal distributions: $\mathcal{Q}^{(1)}(\mu)
\rightarrow \mathcal{Q}^{(0)}(\mu) \rightarrow \mathcal{Q}^{(-1)}(\mu)$ and $\mathcal{R}^{(1)}(\nu) \rightarrow \mathcal{R}^{(0)}(\nu)
\rightarrow \mathcal{R}^{(-1)}(\nu)$. The importance of the quantum-mechanical marginal distributions for $s=0$ in the context of
quantum tomography in discrete phase-space has been stressed by Leonhardt \cite{r16}, where measurements on subensembles of a given
quantum state are necessary in the reconstruction process. 

The Radon transforms represent an important mathematical key for quantum-state reconstruction \cite{r35}. Pursuing this line, Vourdas
\cite{r15} has introduced a wide class of symplectic transformations in Galois quantum systems which allows us to reconstruct the
discrete Wigner function from measurable quantities. Basically, these symplectic transformations consist of Bogoliubov-type unitary
transformations generated by ${\bf J}(\Omega_{1},\Omega_{2},\Omega_{3}) = {\bf M}(\Omega_{3}) {\bf N}(\Omega_{2}) {\bf C}(\Omega_{1})$,
where
\br
{\bf C}(\Omega_{1}) &=& \frac{1}{\sqrt{N}} \sum_{\eta,\xi = -\ell}^{\ell} \exp \lbk - \frac{\im \pi}{N} (1 + \Omega_{1}) \eta \xi \rbk
{\bf S} \lpar \eta , (1 - \Omega_{1}) \xi \rpar \nn \\
{\bf N}(\Omega_{2}) &=& \frac{1}{\sqrt{N}} \sum_{\eta,\xi = -\ell}^{\ell} \exp \lbk \frac{\im \pi}{N} (\Omega_{2} \xi - 2 \eta) \xi \rbk
{\bf S} (\eta , 0) \nn \\
{\bf M}(\Omega_{3}) &=& \frac{1}{\sqrt{N}} \sum_{\eta,\xi = -\ell}^{\ell} \exp \lbk - \frac{\im \pi}{N} (\Omega_{3} \eta + 2 \xi) \eta
\rbk {\bf S} (0 , \xi) \nn
\er
are unitary operators written in terms of the symmetrized Schwinger basis ${\bf S}(\eta,\xi)$, with $\Omega_{1} = \zeta_{4} (1 +
\zeta_{2} \zeta_{3})^{-1}$, $\Omega_{2} = \zeta_{2} \zeta_{4}^{-1} (1+\zeta_{2} \zeta_{3})$, and $\Omega_{3} = \zeta_{3} \zeta_{4} 
(1 + \zeta_{2} \zeta_{3})^{-1}$. Here, the discrete elements of the set $\{ \zeta_{i} \}_{i=1,\ldots,4}$ assume integer values in the
closed interval $[- \ell,\ell ]$, and satisfy the relation $\zeta_{1} \zeta_{4} - \zeta_{2} \zeta_{3} = 1 \, \textrm{mod}(N)$. It is
worth mentioning that this constraint implies in the existence of the inverse elements since $\zeta_{1} = \zeta_{4}^{-1} (1 + \zeta_{2}
\zeta_{3})$. Now, let us initially apply the unitary transformation ${\bf J}(\Omega_{1},\Omega_{2},\Omega_{3})$ on the parametrized
Schwinger basis ${\bf S}^{(s)}(\eta,\xi)$. Thus, after some algebra we obtain
\be
\lb{s3e7}
{\bf J}(\Omega_{1},\Omega_{2},\Omega_{3}) {\bf S}^{(s)}(\eta,\xi) {\bf J}^{\dagger} (\Omega_{1},\Omega_{2},\Omega_{3}) = \lbk
\frac{\mathcal{K}(\zeta_{1} \eta + \zeta_{2} \xi, \zeta_{3} \eta + \zeta_{4} \xi)}{\mathcal{K}(\eta,\xi)} \rbk^{s} {\bf S}^{(s)}
(\zeta_{1} \eta + \zeta_{2} \xi, \zeta_{3} \eta + \zeta_{4} \xi) \; .
\ee
Using this auxiliary result in the calculation of ${\bf J}(\Omega_{1},\Omega_{2},\Omega_{3}) {\bf T}^{(s)}(\mu,\nu) {\bf J}^{\dagger}
(\Omega_{1},\Omega_{2},\Omega_{3})$, we promptly obtain the intermediate expression 
\bd
\frac{1}{\sqrt{N}} \sum_{\eta,\xi = -\ell}^{\ell} \exp \lbk \im \pi \Phi(\eta,\xi;N) - \frac{2 \pi \im}{N} (\eta \mu + \xi \nu) \rbk
\lbk \frac{\mathcal{K}(\zeta_{1} \eta + \zeta_{2} \xi, \zeta_{3} \eta + \zeta_{4} \xi)}{\mathcal{K}(\eta,\xi)} \rbk^{s} {\bf S}^{(s)}
(\zeta_{1} \eta + \zeta_{2} \xi, \zeta_{3} \eta + \zeta_{4} \xi) \; .
\ed
The next step consists in replacing the dummy discrete variables $\eta$ and $\xi$ by $\zeta_{4} \eta^{\prime} - \zeta_{2} \xi^{\prime}$
and $\zeta_{1} \xi^{\prime} - \zeta_{3} \eta^{\prime}$ in the double sum, respectively, with the aim of establishing the compact
expression
\br
\lb{s3e8}
{\bf T}^{(s)}(\mu^{\prime},\nu^{\prime}) &=& \frac{1}{\sqrt{N}} \sum_{\eta^{\prime},\xi^{\prime} = -\ell}^{\ell} \exp \lbk \im \pi
\Phi(\eta^{\prime},\xi^{\prime};N) - \frac{2 \pi \im}{N} (\eta^{\prime} \mu^{\prime} + \xi^{\prime} \nu^{\prime}) \rbk \nn \\
& & \times \lbk \frac{\mathcal{K}(\eta^{\prime},\xi^{\prime})}{\mathcal{K}(\zeta_{4} \eta^{\prime} - \zeta_{2} \xi^{\prime},\zeta_{1}
\xi^{\prime} - \zeta_{3} \eta^{\prime})} \rbk^{s} {\bf S}^{(s)} (\eta^{\prime},\xi^{\prime})
\er
being $\mu^{\prime} = \zeta_{4} \mu - \zeta_{3} \nu$ and $\nu^{\prime} = \zeta_{1} \nu - \zeta_{2} \mu$ the new discrete variables
written as a linear combination of the old ones. In particular, this transformed $\textrm{mod}(N)$ invariant operator basis can be used
to derive the marginal distributions through the standard mathematical procedure
\br
\lb{s3e9}
\mathcal{Q}^{(s)}(\mu;\zeta_{1},\zeta_{3}) &=& \frac{1}{\sqrt{N}} \sum_{\mu^{\prime},\nu^{\prime} = - \ell}^{\ell} 
F^{(s)}(\mu^{\prime},\nu^{\prime}) \, \delta_{\mu, \zeta_{1} \mu^{\prime} + \zeta_{3} \nu^{\prime}}^{[N]} \\
\lb{s3e10}
\mathcal{R}^{(s)}(\nu;\zeta_{2},\zeta_{4}) &=& \frac{1}{\sqrt{N}} \sum_{\mu^{\prime},\nu^{\prime} = - \ell}^{\ell} 
F^{(s)}(\mu^{\prime},\nu^{\prime}) \, \delta_{\nu, \zeta_{2} \mu^{\prime} + \zeta_{4} \nu^{\prime}}^{[N]} \; .
\er
These results characterize the Radon transform in the present context and say that the sum of the parametrized function $F^{(s)}
(\mu^{\prime},\nu^{\prime})$ on specific lines in the $N^{2}$-dimensional phase space represented by the discrete variables
$\mu^{\prime}$ and $\nu^{\prime}$ are equal to the marginal distributions for any value of the parameter $s$ (when $s=0$, the marginal
distributions coincide with probabilities). In terms of the discrete $s$-ordered characteristic function, equations (\ref{s3e9}) and
(\ref{s3e10}) can be written as
\br
\mathcal{Q}^{(s)}(\mu;\zeta_{1},\zeta_{3}) &=& \sum_{\eta = - \ell}^{\ell} \exp \lpar - \frac{2 \pi \im}{N} \eta \mu \rpar \lbk
\frac{\mathcal{K}(\zeta_{1} \eta,\zeta_{3} \eta)}{\mathcal{K}(\eta,0)} \rbk^{s} \Xi^{(s)}(\zeta_{1} \eta, \zeta_{3} \eta) \nn \\
\mathcal{R}^{(s)}(\nu;\zeta_{2},\zeta_{4}) &=& \sum_{\xi = - \ell}^{\ell} \exp \lpar - \frac{2 \pi \im}{N} \xi \nu \rpar \lbk
\frac{\mathcal{K}(\zeta_{2} \xi,\zeta_{4} \xi)}{\mathcal{K}(0,\xi)} \rbk^{s} \Xi^{(s)}(\zeta_{2} \xi, \zeta_{4} \xi) \nn
\er
whose inverse expressions are given by
\br
\lb{s3e11}
\Xi^{(s)}(\zeta_{1} \eta, \zeta_{3} \eta) &=& \frac{1}{N} \lbk \frac{\mathcal{K}(\eta,0)}{\mathcal{K}(\zeta_{1} \eta, \zeta_{3} \eta)}
\rbk^{s} \sum_{\mu = -\ell}^{\ell} \exp \lpar \frac{2 \pi \im}{N} \mu \eta \rpar \mathcal{Q}^{(s)}(\mu;\zeta_{1},\zeta_{3}) \\
\lb{s3e12}
\Xi^{(s)}(\zeta_{2} \xi, \zeta_{4} \xi) &=& \frac{1}{N} \lbk \frac{\mathcal{K}(0,\xi)}{\mathcal{K}(\zeta_{2} \xi, \zeta_{4} \xi)}
\rbk^{s} \sum_{\nu = -\ell}^{\ell} \exp \lpar \frac{2 \pi \im}{N} \nu \xi \rpar \mathcal{R}^{(s)}(\nu;\zeta_{2},\zeta_{4}) \; .
\er
Note that equations (\ref{s3e11}) and (\ref{s3e12}) establish a link between measurable quantities (rhs) and discrete $s$-ordered
characteristic functions (lhs); moreover, they can be used to construct, for instance, the quasiprobability distribution functions in
finite-dimensional spaces. In summary, we have established a set of important theoretical results which constitute a discrete version
of that obtained by Vogel and Risken \cite{r36} for the continuous case.

From the theoretical point of view, the ideas of quantum computation can nowadays be used for illuminating some fundamental processes
in quantum mechanics \cite{r37}. In this sense, Paz and co-workers \cite{r17} have shown that tomography and spectroscopy are dual
forms of the same quantum computation (represented by a `scattering' circuit), since the state of a quantum system can be modeled on a
quantum computer. Furthermore, using different versions of programmable gate arrays, the authors have been capable not only of
evaluating the expectation value of any operator acting on an $N$-dimensional space of states, but also of measuring other probability
distribution functions (e.g., Husimi and Kirkwood functions) in a discrete phase-space. Here, we employ a slightly modified version of
the scattering circuit to measure the discrete Wigner function $\mathcal{W}(\mu,\nu)$. Basically, we modify this circuit by inserting a
controlled-${\bf U}$ operation between the Hadamard gates, with ${\bf U} = \sqrt{N} {\bf S}(\eta,\xi)$ acting on a quantum system
described by some unknown density operator $\ro$, and also a controlled Fourier transform (FT) after the second Hadamard gate. This is
illustrated in figure 1, where a set of measurements on the polarizations along the $z$ and $y$ axes of the ancillary qubit $| 0 \rg$
yield the expectation values $\lg \sig_{z} \rg = \sqrt{N} \re [ \mathcal{W}(\mu,\nu) ]$ and $\lg \sig_{y} \rg = \sqrt{N} \ima [
\mathcal{W}(\mu,\nu) ]$, respectively. In the absence of the controlled-FT operation, these measurements lead us to obtain the
characteristic function $\Xi^{(s)}(\eta,\xi)$ for $s=0$, namely, $\lg \sig_{z} \rg = \sqrt{N} \re [ \Xi^{(0)}(\eta,\xi) ]$ and $\lg
\sig_{y} \rg = \sqrt{N} \ima [ \Xi^{(0)}(\eta,\xi) ]$. However, to construct the discrete Husimi function, some modifications must be
included in the primary circuit (see reference \cite{r17} for more details) or the link established by equation (\ref{s2e11}) between
the Wigner and Husimi functions should be employed. Both situations deserve a detailed theoretical investigation since their operational
costs can be prohibitive from the experimental point of view. Next, we will present a phase-space description of the process inherent to
quantum teleportation for a system with an $N$-dimensional space of states. 
\begin{figure}[!t]
\centering
\begin{minipage}[b]{0.70\linewidth}
\includegraphics[width=1.00\textwidth]{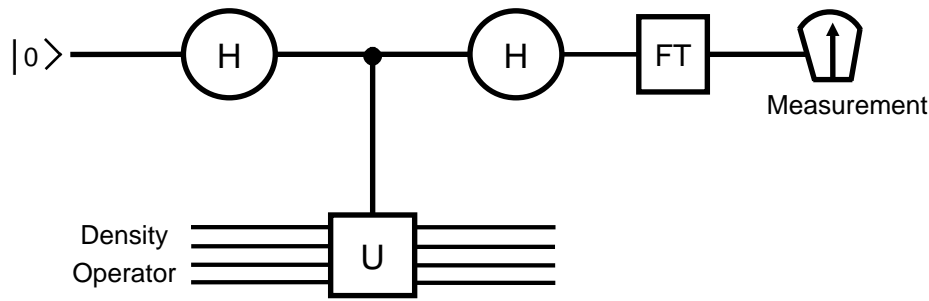}
\end{minipage}
\caption{Slightly modified version of the ``scattering circuit" used to evaluate the real and imaginary parts of the expectation value 
$\tr ({\bf U} \ro)$ for a unitary operator ${\bf U}$, where $| 0 \rg$ represents the ancillary qubit state which acts as a probe
particle in a scattering experiment, and $\textrm{\bf H}$ denotes a Hadamard transform. In particular, for a controlled-${\bf U}$
operation given by ${\bf U} = \sqrt{N} {\bf S}(\eta,\xi)$, the measurements of the ancillary qubit polarizations along the $z$ and
$y$ axes allow us to construct the discrete Wigner function $\mathcal{W}(\mu,\nu) = \tr [ {\bf T}^{(0)}(\mu,\nu) \ro ]$ (discrete
characteristic function $\Xi^{(0)}(\eta,\xi) = \tr [ {\bf S}^{(0)}(\eta,\xi) \ro ]$) in the presence (absence) of the controlled-FT
operation.}
\end{figure}

\subsection{Discrete phase-space representation of quantum teleportation}

In the last years, great advance has been reached in the quantum teleportation arena. In particular, we observe that: (i) different
theoretical schemes for teleportation of quantum states involving continuous and discrete variables have been proposed and investigated
in the literature \cite{r18,r19,r20,r21,r38}, and (ii) its experimental feasibility has been demonstrated in simple systems through
pairs of entangled photons produced by the process of parametric down-conversion \cite{r39}. Moreover, the essential resource in both
theoretical and experimental approaches is directly associated with the concept of entanglement, which naturally appears in quantum
mechanics when the superposition principle is applied to composite systems. An immediate consequence of this important effect has its
origin in the theory of quantum measurement \cite{r40}, since the entangled state of the multipartite system can reveal information
about its constituent parts. 

Recently, the quasiprobability distribution functions have represented important tools in the phase-space description of the quantum
teleportation process for a system with an $N$-dimensional space of states. For instance, Koniorczyk {\em et al} \cite{r19} have 
presented a unified approach to quantum teleportation in arbitrary dimensions based on the Wigner-function formalism, where the finite-
and infinite-dimensional cases can be treated in a conceptually uniform way. Paz \cite{r20} has extended the results obtained by
Koniorczyk {\em et al} to the case where the space of states has arbitrary dimensionality. To this end, the author has used a different
definition for the discrete Wigner function which permits us to analyze situations where entanglement among subsystems of arbitrary
dimensionality is an important issue. Here, we use the new ${\rm mod}(N)$ invariant operator basis ${\bf T}^{(s)}(\mu,\nu)$ in order to
obtain a discrete phase-space representation of quantum teleportation which permits us to extend the results reached by Paz in the
discrete Wigner-function context for any discrete quasiprobability distribution functions.

\subsubsection{Generalized Bell states}

The generalized Bell states were first introduced by Bennett {\em et al} \cite{r18} in the study of quantum teleportation for systems
with $N > 2$ orthogonal states. Basically, these states can be defined as $| \Psi_{\om_{1},\om_{2}} \rg = {\bf V}_{1}^{\om_{1}} \otimes
{\bf U}_{2}^{-\om_{2}} | \Psi_{0,0} \rg$, where
\bd
| \Psi_{0,0} \rg = \frac{1}{\sqrt{N}} \sum_{\eps = -\ell}^{\ell} | v_{\eps} \rg_{1} \otimes | v_{\eps} \rg_{2}
\ed
represents the pure state maximally entangled for a bipartite system (in this case, the reduced density matrix of each constituent
part is equal to $(1/N) {\bf 1}_{i}$ for $i=1,2$), being $\{ | v_{\alpha} \rg_{i} \}_{\alpha = - \ell,\ldots,\ell}$ and $\{ | u_{\beta}
\rg_{j} \}_{\beta = - \ell,\ldots,\ell}$ the eigenstates of the Schwinger unitary operators ${\bf V}_{i}$ and ${\bf U}_{j}$, 
respectively. Furthermore, the generalized Bell states satisfy the following properties:
\br
& \mbox{(i)} & \lg \Psi_{\om_{1},\om_{2}} | \Psi_{\om_{1}^{\prime},\om_{2}^{\prime}} \rg = \delta_{\om_{1}^{\prime},\om_{1}}^{[N]}
\delta_{\om_{2}^{\prime},\om_{2}}^{[N]} \qquad \mbox{(orthogonality relation)} \nn \\
& \mbox{(ii)} & \sum_{\om_{1},\om_{2} = - \ell}^{\ell} | \Psi_{\om_{1},\om_{2}} \rg \lg \Psi_{\om_{1},\om_{2}} | = {\bf 1}_{1} \otimes
{\bf 1}_{2} \qquad \mbox{(identity relation)} \nn \\
& \mbox{(iii)} & {\bf U}_{+} | \Psi_{\om_{1},\om_{2}} \rg = {\bf U}_{1} \otimes {\bf U}_{2} | \Psi_{\om_{1},\om_{2}} \rg = \exp [ -
( 2 \pi \im /N ) \, \om_{1} ] | \Psi_{\om_{1},\om_{2}} \rg \nn \\
& & {\bf V}_{-} | \Psi_{\om_{1},\om_{2}} \rg = {\bf V}_{1} \otimes {\bf V}_{2}^{-1} | \Psi_{\om_{1},\om_{2}} \rg = \exp [ (2 \pi \im /N)
\, \om_{2} ]  | \Psi_{\om_{1},\om_{2}} \rg \; . \nn 
\er
Note that ${\bf U}_{+}$ displaces both systems in coordinate by the same amount $\om_{1}$, while ${\bf V}_{-}$ displaces them in
momentum by the quantity $\om_{2}$ in the opposite direction. In addition, as $\{ | \Psi_{\om_{1},\om_{2}} \rg \}_{\om_{1},\om_{2} = -
\ell,\ldots,\ell}$ are common eigenstates of ${\bf U}_{+}$ and ${\bf V}_{-}$, such states can be interpreted as corresponding to the
eigenstates of the total momentum and relative coordinate operators \cite{r19,r20}; indeed, these states are the discrete version of
the continuous ones used by Einstein, Podolsky, and Rosen \cite{r41}. Thus, the generalized Bell measurements will be characterized in
our context by the set of diagonal projection operators $\{ | \Psi_{\om_{1},\om_{2}} \rg \lg \Psi_{\om_{1},\om_{2}} | \}_{\om_{1},
\om_{2}  = - \ell,\ldots,\ell}$.

Now, let us establish some further results related to the generalized Bell states and their discrete phase-space representation. The
first one corresponds to the mapping of $| \Psi_{\om_{1},\om_{2}} \rg \lg \Psi_{\om_{1}^{\prime},\om_{2}^{\prime}} |$ in terms of the
basis $\{ {\bf T}^{(s_{i})}_{i}(\mu_{i},\nu_{i}) \}_{\mu_{i},\nu_{i} = - \ell,\ldots,\ell}$ for each subsystem, i.e.,
\be
\lb{s3e13}
| \Psi_{\om_{1},\om_{2}} \rg \lg \Psi_{\om_{1}^{\prime},\om_{2}^{\prime}} | = \frac{1}{N^{2}} \sum_{\mu_{1},\nu_{1},\mu_{2},\nu_{2} = 
- \ell}^{\ell} \Upsilon^{(-s_{1},-s_{2})} (\om_{1},\om_{2},\om_{1}^{\prime},\om_{2}^{\prime} | \mu_{1},\nu_{1},\mu_{2},\nu_{2}) \, 
{\bf T}_{1}^{(s_{1})}(\mu_{1},\nu_{1}) \otimes {\bf T}_{2}^{(s_{2})}(\mu_{2},\nu_{2})
\ee
with the coefficients of the expansion given by
\bd
\Upsilon^{(-s_{1},-s_{2})} (\om_{1},\om_{2},\om_{1}^{\prime},\om_{2}^{\prime} | \mu_{1},\nu_{1},\mu_{2},\nu_{2}) = \tr \lbk 
{\bf T}_{1}^{(-s_{1})} (\mu_{1},\nu_{1}) \otimes {\bf T}_{2}^{(-s_{2})}(\mu_{2},\nu_{2}) | \Psi_{\om_{1},\om_{2}} \rg \lg
\Psi_{\om_{1}^{\prime},\om_{2}^{\prime}} | \rbk \; .
\ed
Consequently, the second one refers to the inverse mapping of (\ref{s3e13}), which can be directly reached with the help of property
(ii) as follows:
\be
\lb{s3e14}
{\bf T}_{1}^{(s_{1})} (\mu_{1},\nu_{1}) \otimes {\bf T}_{2}^{(s_{2})}(\mu_{2},\nu_{2}) = \sum_{\om_{1},\om_{2},\om_{1}^{\prime},
\om_{2}^{\prime} = - \ell}^{\ell} \Theta^{(s_{1},s_{2})}(\mu_{1},\nu_{1},\mu_{2},\nu_{2} | \om_{1},\om_{2},\om_{1}^{\prime},
\om_{2}^{\prime}) \, | \Psi_{\om_{1},\om_{2}} \rg \lg \Psi_{\om_{1}^{\prime},\om_{2}^{\prime}} |
\ee
being
\bd
\Theta^{(s_{1},s_{2})} (\mu_{1},\nu_{1},\mu_{2},\nu_{2} | \om_{1},\om_{2},\om_{1}^{\prime},\om_{2}^{\prime}) = \tr \lbk 
{\bf T}_{1}^{(s_{1})} (\mu_{1},\nu_{1}) \otimes {\bf T}_{2}^{(s_{2})}(\mu_{2},\nu_{2}) | \Psi_{\om_{1}^{\prime},\om_{2}^{\prime}}
\rg \lg \Psi_{\om_{1},\om_{2}} | \rbk \; .
\ed
It is worth mentioning that a general connection between the coefficients of both expansions (\ref{s3e13}) and (\ref{s3e14}) can also
be promptly established for any values of $\{ s_{1},s_{2} \} \in \mathbb{R}$,
\br
\Upsilon^{(-s_{1},-s_{2})} (\om_{1},\om_{2},\om_{1}^{\prime},\om_{2}^{\prime} | \mu_{1},\nu_{1},\mu_{2},\nu_{2}) &=&
\Theta^{(-s_{1},-s_{2})} (\mu_{1},\nu_{1},\mu_{2},\nu_{2} | \om_{1}^{\prime},\om_{2}^{\prime},\om_{1},\om_{2}) \nn \\
&=& \lbk \Theta^{(-s_{1},-s_{2})} (\mu_{1},\nu_{1},\mu_{2},\nu_{2} | \om_{1},\om_{2},\om_{1}^{\prime},\om_{2}^{\prime}) 
\rbk^{\ast} \; . \nn
\er
The analytical expression of these coefficients will be omitted here due to its apparent irrelevance in the phase-space description of
the quantum teleportation process. 

However, some useful results derived from these coefficients deserve to be mentioned and discussed in detail. For instance, equation
(\ref{s3e14}) allows us to calculate the parametrized function
\be
\lb{s3e15}
F^{(s_{1},s_{2})}_{\om_{1},\om_{2}}(\mu_{1},\nu_{1},\mu_{2},\nu_{2}) = \tr \lbk {\bf T}_{1}^{(s_{1})} (\mu_{1},\nu_{1}) \otimes 
{\bf T}_{2}^{(s_{2})}(\mu_{2},\nu_{2}) | \Psi_{\om_{1},\om_{2}} \rg \lg \Psi_{\om_{1},\om_{2}} | \rbk
\ee
which coincides with $\Theta$ for particular values of $\om_{i}$ and $\om_{i}^{\prime}$. In this situation, the analytical expression
\bd
F^{(s_{1},s_{2})}_{\om_{1},\om_{2}}(\mu_{1},\nu_{1},\mu_{2},\nu_{2}) = \frac{1}{N^{2}} \sum_{\eta,\xi = - \ell}^{\ell} \exp \lbr
\frac{2 \pi \im}{N} \lbk \eta (\mu_{1} + \mu_{2} + \om_{1}) + \xi (\nu_{1} - \nu_{2} - \om_{2}) \rbk \rbr \lbk \mathcal{K}(\eta,\xi)
\rbk^{-(s_{1} + s_{2})}
\ed
can be reduced to the following discrete quasiprobability distribution functions:
\br
& \bullet & s_{1} = s_{2} = 0 \; \; \mbox{(Wigner function)} \qquad \mathcal{W}_{\om_{1},\om_{2}}(\mu_{1},\nu_{1},\mu_{2},\nu_{2}) = 
\delta_{\om_{1},-(\mu_{1} + \mu_{2})}^{[N]} \delta_{\om_{2},\nu_{1} - \nu_{2}}^{[N]} \nn \\
& \bullet & s_{1} = s_{2} = -1 \; \; \mbox{(Husimi function)} \qquad \! \! \! \mathcal{H}_{\om_{1},\om_{2}} (\mu_{1},\nu_{1},\mu_{2},
\nu_{2}) = N^{-1} | \mathcal{K}(\mu_{1} + \mu_{2} + \om_{1} , \nu_{1} - \nu_{2} - \om_{2}) |^{2} \; . \nn
\er
To measure the discrete Wigner function associated with the generalized Bell states, some minor modifications should be implemented in
the scattering circuit (see figure 1): the first one concerns to the controlled-${\bf U}$ operation between the Hadamard gates, since
it must be replaced by ${\bf U} = [ \sqrt{N} {\bf S}_{1}(\eta_{1},\xi_{1}) ] \otimes [ \sqrt{N} {\bf S}_{2}(\eta_{2},\xi_{2}) ]$ in
order to process operations for bipartite systems; while the second one consists in preparing the input density operator in the
generalized Bell states, namely $\ro = | \Psi_{\om_{1},\om_{2}} \rg \lg \Psi_{\om_{1},\om_{2}} |$. This procedure leads us to obtain
the expectation value $\lg \sig_{z} \rg = N \mathcal{W}_{\om_{1},\om_{2}}(\mu_{1},\nu_{1},\mu_{2},\nu_{2})$ through a set of measurements
on the polarization along the $z$-axis of the ancillary qubit. Furthermore, these minor modifications on the scattering circuit can also
be used to measure any discrete Wigner function associated with a general bipartite system.

\subsubsection{Quantum teleportation}

Basically, the quantum teleportation process consists in a sequence of events that allows us to transfer the quantum state of a particle
onto another particle through an essential feature of quantum mechanics: entanglement \cite{r18,r39}. In this sense, let us introduce a
tripartite system described by $\ro = \ro_{1} \otimes ( | \Psi_{0,0} \rg \lg \Psi_{0,0} | )_{23}$, where subsystems 2 and 3 were
initially prepared in one of the Bell states. The plan is to teleport the initial state of subsystem 1 through the protocol established
in \cite{r20}.
\begin{enumerate}
\item We initiate the protocol considering the density operator associated with the tripartite system written in terms of the new basis
$\{ {\bf T}^{(s_{i})}_{i}(\mu_{i},\nu_{i}) \}_{\mu_{i},\nu_{i} = - \ell,\ldots,\ell}$ for each subsystem $i=1,2,3$ as follows:
\bd
\ro = \frac{1}{N^{3}} \sum_{\mu_{1},\nu_{1},\mu_{2},\nu_{2},\mu_{3},\nu_{3} = - \ell}^{\ell} \, F_{1}^{(-s_{1})}(\mu_{1},\nu_{1})
F_{23}^{(-s_{2},-s_{3})}(\mu_{2},\nu_{2},\mu_{3},\nu_{3}) {\bf T}_{1}^{(s_{1})}(\mu_{1},\nu_{1}) \otimes {\bf T}_{2}^{(s_{2})}
(\mu_{2},\nu_{2}) \otimes {\bf T}_{3}^{(s_{3})}(\mu_{3},\nu_{3})
\ed
where
\br
F_{1}^{(-s_{1})}(\mu_{1},\nu_{1}) &=& \tr_{1} \lbk {\bf T}_{1}^{(-s_{1})}(\mu_{1},\nu_{1}) \ro_{1} \rbk \nn \\
F_{23}^{(-s_{2},-s_{3})}(\mu_{2},\nu_{2},\mu_{3},\nu_{3}) &=& \tr_{23} \lbk {\bf T}_{2}^{(-s_{2})}(\mu_{2},\nu_{2}) \otimes
{\bf T}_{3}^{(-s_{3})}(\mu_{3},\nu_{3}) \, ( | \Psi_{0,0} \rg \lg \Psi_{0,0} | )_{23} \rbk \; . \nn
\er
Next, we perform a measurement on subsystems 1 and 2 that projects them into the Bell states (this procedure corresponds to a collective
measurement which determines the total momentum and relative coordinate for composite subsystem 1-2). For convenience, before the
generalized Bell measurement, let us express the phase-space operators ${\bf T}_{1}^{(s_{1})}(\mu_{1},\nu_{1}) \otimes 
{\bf T}_{2}^{(s_{2})}(\mu_{2},\nu_{2})$ according to equation (\ref{s3e14}),
\br
\ro &=& \frac{1}{N^{3}} \sum_{\mu_{1},\nu_{1},\mu_{2},\nu_{2},\mu_{3},\nu_{3} = - \ell}^{\ell} F_{1}^{(-s_{1})}(\mu_{1},\nu_{1})
F_{23}^{(-s_{2},-s_{3})}(\mu_{2},\nu_{2},\mu_{3},\nu_{3}) \nn \\
& & \times \sum_{\om_{1},\om_{2},\om_{1}^{\prime},\om_{2}^{\prime} = - \ell}^{\ell} \Theta^{(s_{1},s_{2})} (\mu_{1},\nu_{1}, \mu_{2},
\nu_{2} | \om_{1},\om_{2},\om_{1}^{\prime},\om_{2}^{\prime}) \, ( | \Psi_{\om_{1},\om_{2}} \rg \lg \Psi_{\om_{1}^{\prime},
\om_{2}^{\prime}} | )_{12} \otimes {\bf T}_{3}^{(s_{3})}(\mu_{3},\nu_{3}) \; . \nn
\er
Thus, after the measurement on the first two subsystems, only the terms with $\om_{1} = \om_{1}^{\prime} = \alf$ and $\om_{2} =
\om_{2}^{\prime} = \bet$ survive. Consequently, a reduced density operator for the third subsystem can be promptly obtained,
\be
\lb{s3e16}
\ro_{3R} = \frac{1}{N} \sum_{\mu_{3},\nu_{3} = - \ell}^{\ell} \Lambda_{\alf,\bet}^{(-s_{1},-s_{3})}(\mu_{3},\nu_{3}) \,
{\bf T}_{3}^{(s_{3})}(\mu_{3},\nu_{3}) \; ,
\ee
which does not depend on the complex parameter $s_{2}$. Here, the coefficients are given by
\bd
\Lambda_{\alf,\bet}^{(-s_{1},-s_{3})}(\mu_{3},\nu_{3}) = \sum_{\mu_{1},\nu_{1}=- \ell}^{\ell} \mathfrak{R}_{\alf,\bet}^{(s_{3}-s_{1})}
(\mu_{1},\nu_{1},\mu_{3},\nu_{3}) F_{1}^{(-s_{1})}(\mu_{1},\nu_{1})
\ed
with
\bd
\mathfrak{R}_{\alf,\bet}^{(s_{3}-s_{1})}(\mu_{1},\nu_{1},\mu_{3},\nu_{3}) = \frac{1}{N^{2}} \sum_{\eta,\xi = - \ell}^{\ell} \exp \lbr 
\frac{2 \pi \im}{N} \lbk \eta (\mu_{1} - \mu_{3} + \alf) - \xi (\nu_{1} - \nu_{3} - \bet) \rbk \rbr \lbk \mathcal{K}(\eta,\xi) 
\rbk^{s_{3}-s_{1}} \; .
\ed
Note that $\Lambda_{\alf,\bet}^{(-s_{1},-s_{3})}(\mu_{3},\nu_{3})$ simply tells us how to construct, independently of parameters
$s_{1}$ and $s_{3}$, the final parametrized function for the third subsystem from the initial parametrized function of the first one.

\item Now, let us analyze the particular case $s_{1}=s_{3}=s$. In this situation, equation (\ref{s3e16}) assumes the simplified form
\be
\lb{s3e17}
\ro_{3R} = \frac{1}{N} \sum_{\mu_{3},\nu_{3} = - \ell}^{\ell} F_{1}^{(-s)}(\mu_{3} - \alf,\nu_{3} + \bet) \, {\bf T}_{3}^{(s)}
(\mu_{3},\nu_{3})
\ee
where the coefficients $F_{1}^{(-s)}(\mu_{3} - \alf,\nu_{3} + \bet)$ play a central role in the phase-space description of the quantum
teleportation process. In fact, they allow us to conclude that, after the generalized Bell measurement, the third subsystem has a
parametrized function which is displaced in phase-space by an amount $(-\alf,\bet)$ with respect to the initial state of the first
subsystem, namely $F_{3R}^{(-s)}(\mu_{3},\nu_{3}) = F_{1}^{(-s)}(\mu_{3}-\alf , \nu_{3}+\bet)$. Therefore, the recovery operation
basically depends on the calibration process of the generalized Bell measurements performed on the first two subsystems: for instance,
when $\alf=\bet=0$, we reach a complete recovery operation.
\end{enumerate}

In short, we have presented a quantum teleportation protocol that leads us to obtain a phase-space description of this process for
any discrete quasiprobability distribution functions associated with physical systems described by $N$-dimensional space of states.

\section{Conclusions}

In this paper we have employed the new mod$(N)$ invariant operator basis $\{ {\bf T}^{(s)}(\mu,\nu) \}_{\mu,\nu = - \ell,\ldots,\ell}$
recently proposed in \cite{r26}, with the aim of obtaining $s$-parametrized phase-space functions which are responsible for the mapping
of bounded operators, acting on a finite-dimensional Hilbert space, on their discrete representatives in an $N^{2}$-dimensional phase
space. In fact, we have established a set of important formal results that allows us to reach a discrete analog of the continuous one
developed by Cahill and Glauber \cite{r1}. As a consequence, the discrete Glauber-Sudarshan $(s=1)$, Wigner $(s=0)$, and Husimi $(s=-1)$
functions emerge from this formalism as specific cases of $s$-parametrized phase-space functions describing density operators associated
with physical systems whose space of states has a finite dimension. In addition, we have also established a hierarchical order among 
them that consists of a well-defined smoothing process where, in particular, the kernel $\mathcal{K}(\eta,\xi)$ performs a central role.
Next, we have applied our formalism to the context of quantum information processing, quantum tomography, and quantum teleportation in
order to obtain a phase-space description of some topics related to unitary depolarizers, discrete Radon transforms, and generalized
Bell states. Indeed such descriptions have allowed us to attain new important results, within which some deserve to be mentioned:
(i) we have shown that the symmetrized version of the Schwinger operator basis $\{ {\bf S}(\eta,\xi) \}_{\eta,\xi = - \ell,\ldots,\ell}$
can be considered a unitary depolarizer; (ii) we have also established a link between measurable quantities and discrete $s$-ordered
characteristic functions with the help of Radon transforms, which can be used to construct any quasiprobability distribution functions
in finite-dimensional spaces; and finally, (iii) we have presented a quantum teleportation protocol that leads us to obtain a
generalized phase-space description of this important process in physics. It is worth mentioning that the mathematical formalism
developed here opens new possibilities of future investigations in similar physical systems \cite{r42}; or in the study of dissipative
systems, where the decoherence effect has a central role in the quantum information processing (e.g., see reference \cite{r23}). These
considerations are under current research and will be published elsewhere.
 
\section*{Acknowledgments}

This work has been supported by Funda\c{c}\~{a}o de Amparo \`{a} Pesquisa do Estado de S\~{a}o Paulo (FAPESP), Brazil, project nos.
01/11209-0 (MAM), 03/13488-0 (MR) and 00/15084-5 (MAM and MR). DG acknowledges partial financial support from the Conselho Nacional de
Desenvolvimento Cient\'{\i}fico e Tecnol\'{o}gico (CNPq), Brazil.


\end{document}